\documentclass[prd,aps,noshowpacs,nofootinbib, 11 pt]{revtex4}
\usepackage{amsmath,graphicx,color,epsfig}

\begin{document}

\title{Lepton polarization asymmetry and forward backward asymmetry in
exclusive $B\rightarrow K_{1}\tau ^{+}\tau ^{-}$ decay in universal extra
dimension scenario}

\author{Asif Saddique$^{1}$} \author{M. Jamil\ Aslam$^{2,3}$} \author{Cai-Dian L\"{u}$^{2}$}

\vspace*{1.0cm}

\affiliation{$^{1}$Department of Physics, Quaid-i-Azam University, Islamabad\\
and National Centre for Physics, Islamabad, Pakistan. \\
$^{2}$Institute of High Energy Physic, P. O.\ Box 918(4), Beijing,
100049, P.\ R. China. \\
$^{3}$COMSATS Institute of Information Technology Islamabad,
Pakistan}

\begin{abstract}
Decay rate, forward-backward asymmetry and polarization asymmetries of final
state leptons in $B\rightarrow K_{1}\tau ^{+}\tau ^{-}$, where $K_{1}$ is
the axial vector meson, are calculated in Standard Model and in the
universal extra dimension (UED) model. The sensitivity of the observables on
the compactification radius $R$, the only unknown paramter in UED\ model, is
studied. Finally, the helicity fractions of the final state $K_{1}$ are
calculated and their dependence on the compactification radius is discussed.
This analysis of helicity fraction is briefly extended to $B\rightarrow
K^{\ast }\ell ^{+}\ell ^{-}$ $\left( \ell =e,\mu \right) $ and compared with
the other approaches exist in the literature$.$
\end{abstract}

 \maketitle

\section{Introduction}

It is generally believed that Standard Model (SM) of particle physics is one
of the most successful theory of the second half of previous century in
explaining the observed data so far, but no one can say that it is the end
of physics. Intensive search for physics beyond SM is now being performed in
various areas of particle physic which is expected to get the direct
evidence at high energy colliders such as the Large Hadron Collider (LHC).
During the last years there has been an increased interest in models with
extra dimensions, since they solve the hierarchy problem and can provide the
unified framework of gravity and other interactions together with a
connection with the string theory \cite{01}. Among them a special role play
the ones with universal extra dimensions (UED) as in these models all SM
fields are allowed to propagate in all available dimensions. The economy of
UED models is that there is only one new free parameter in addition to SM,
the radius $R$ of the compactified extra dimension. Now above the
compactification scale $1/R$ \ a given UED model become a higher dimensional
field theory whose equivalent description in four dimensions consists of SM
fields, the towers of their Kaluza-Klein (KK) partners and additional towers
of KK modes having no partner in SM. Appelquist, Cheng and Dobrescu (ACD)
model \cite{02}, with one extra universal dimension is the simplest model of
this type. In this model the only additional free parameter relative to SM
is the compactification scale $1/R$. Thus, all the masses of the KK
particles and their interactions with SM particles and also among themselves
are described in terms of $1/R$ and the parameters of SM \cite{03}.

The most profound property of ACD model is the conservation of KK parity
which implies the absences of tree level contribution of KK states to the
low energy processes taking place at scale $\mu \ll 1/R$. This brings
interest towards the flavor-changing-neutral-current (FCNC) transitions $%
b\rightarrow s$, as these are not allowed at tree level but are induced by
the Glashow-Iliopoulos-Miani (GIM) amplitudes \cite{03a} at the loop level
in the SM and hence the one loop contribution due to KK modes to them could
in principle be important. These processes are used to constrain the mass
and couplings of KK\ states, i.e. the compactification parameter $1/R$ \cite%
{04}.

Buras and collaborators have computed the effective Hamiltonian of several
FCNC processes in ACD model, particularly in $b$ sector, namely $B_{s,d}$
mixing and $b\rightarrow s$ transition such as $b\rightarrow s\gamma $ and $%
b\rightarrow s\ell ^{+}\ell ^{-}$ \cite{03}. The implications of physics
with UED are being examined with the data from accelerator experiments, for
example, from Tevatron experiments the bound on the inverse of
compactification radius is found to be about $1/R\geq 300$ GeV \cite{05}.
Exclusive $B\rightarrow K\left( K^{\ast }\right) \ell ^{+}\ell ^{-}$, $%
B\rightarrow K\left( K^{\ast }\right) \nu \bar{\nu}$ and $B\rightarrow
K^{\ast }\gamma $ decays are analyzed in ACD model and it was shown that the
uncertainties connected with hadronic matrix elements does not mask the
sensitivity to the compactification parameter, and the current data on the
decay rates of $B\rightarrow K^{\ast }\gamma $ and $B\rightarrow K^{\ast
}\ell ^{+}\ell ^{-}$ $\left( \ell =e,\mu \right) $ can provide a similar
bound to the inverse compactification radius: $1/R\geq 300-400$ GeV \cite{06}%
. In addition to these the decay modes $B\rightarrow K_{1}\ell ^{+}\ell ^{-}$
$\left( \ell =e,\mu \right) $, $B\rightarrow \phi \ell ^{+}\ell ^{-}$, $%
B\rightarrow \gamma \ell ^{+}\ell ^{-}$ \ and $\Lambda _{b}\rightarrow
\Lambda \ell ^{+}\ell ^{-}$ have also been considered, with the possibility
of observing such processes at hadron colliders \cite{07, 08, 09}.

Colangelo \textit{et al}. have also considered another set of observables in
FCNC\ transitions, namely those of the inclusive $B\rightarrow X_{s}+leptons$
and exclusive $B\rightarrow K\left( K^{\ast }\right) +leptons$ decay modes,
where the leptons are $\tau ^{+}\tau ^{-}$ \cite{09a}. There is no
experimental data on these days as yet, however as first noticed in \cite{10}%
, these processes are of great interest due to the possibility of measuring
lepton polarization asymmetries which are sensitive to the structure of the
interactions, so that they can be used to test the SM and its extensions.
They analyzed the $\tau ^{-}$ polarization asymmetries in single universal
extra dimension model both for inclusive and exclusive semileptonic $B$
meson decays. Besides this, they investigated another observable, the
fraction of longitudinal $K^{\ast }$ polarization in $B\rightarrow K^{\ast
}\ell ^{+}\ell ^{-}$, for which a new measurement in two bins of momentum
transfer to the lepton pair is available in case of $\ell =e,\mu $. They
studied the dependence of this quantity on the compactification parameter,
for $B\rightarrow K^{\ast }\tau ^{+}\tau ^{-}$ and in the case of light
leptons, together with the fraction of $K^{\ast }$ polarization in the same
modes, and discussed the possibility to constrain the universal extra
dimension scenario.

In this work, we will study the spin effects on $B\rightarrow K_{1}\tau
^{+}\tau ^{-}$ in ACD model using the framework of $B\rightarrow K^{\ast
}\tau ^{+}\tau ^{-}$ described by Colangelo \textit{et al}. \cite{09a}. We
investigate the branching ratio, forward backward and polarization
asymmetries for the final state $\tau ^{-}$. Although the sensitivity of
branching ratio and forward backward asymmetry on the extra dimension is
mild but still we believe that together with the $\tau ^{-}$ lepton
polarization asymmetries, these can be used to provide additional
constraints on the comactification parameter. In extension to this, we have
also discussed, the fraction of longitudinal $K^{\ast }$ polarization in $%
B\rightarrow K^{\ast }\ell ^{+}\ell ^{-}$, for which new measurements in two
bins of momentum transfer to the lepton pair is available in case of $\ell
=e $, $\mu $ and compared them with the other approaches existed already in
the literature \cite{09a}. Finally, we have used the same method to
calculate the helicity fractions of $K_{1}$ in $B\rightarrow K_{1}\ell
^{+}\ell ^{-}$ both in SM and in ACD model. We hope that these fractions put
another useful constraints on the universal extra dimension scenario.

The paper is organized as follows. In Section 2 we present the effective
Hamiltonian for $B\rightarrow K_{1}\ell ^{+}\ell ^{-}$ in ACD model. In
section 3, we will calculate the decay rate and forward backward asymmetry
for $B\rightarrow K_{1}\tau ^{+}\tau ^{-}$. Section 4 and 5 deals with the
study of polarization asymmetries of final state $\tau ^{-}$ and the
helicity fractions of final state $K_{1}$ meson, respectively. We will
summarize our results at the last section.

\section{Effective Hamiltonian}

At quark level the decay $B\rightarrow K_{1}\ell ^{+}\ell ^{-}$ is same like
$B\rightarrow K^{\ast }\ell ^{+}\ell ^{-}$ as discussed by Ali \textit{et al}%
.\cite{11}, i.e. $b\rightarrow s\ell ^{+}\ell ^{-}$ and it can be described
by effective Hamiltonian obtained by integrating out the top quark and $%
W^{\pm }$ bosons
\begin{equation}
H_{eff}=-4\frac{G_{F}}{\sqrt{2}}V_{tb}V_{ts}^{\ast }\sum_{i=1}^{10}C_{i}(\mu
)O_{i}(\mu )  \label{2.1}
\end{equation}%
where $O_{i}$'$s$ are four local quark operators and $C_{i}$ are Wilson
coefficients calculated in Naive dimensional regularization (NDR) scheme %
\cite{12}.

One can write the above Hamiltonian in the following free quark decay
amplitude
\begin{eqnarray}
\mathcal{M}(b &\rightarrow &s\ell ^{+}\ell ^{-})=\frac{G_{F}\alpha }{\sqrt{2}%
\pi }V_{tb}V_{ts}^{\ast }\left\{
\begin{array}{c}
C_{9}^{eff}\left[ \bar{s}\gamma _{\mu }Lb\right] \left[ \bar{\ell}\gamma
^{\mu }\ell \right] \\
+C_{10}\left[ \bar{s}\gamma _{\mu }Lb\right] \left[ \bar{\ell}\gamma ^{\mu
}\gamma ^{5}\ell \right] \\
-2\hat{m}_{b}C_{7}^{eff}\left[ \bar{s}i\sigma _{\mu \nu }\frac{\hat{q}^{\nu }%
}{\hat{s}}Rb\right] \left[ \bar{\ell}\gamma ^{\mu }\ell \right]%
\end{array}%
\right\} ,  \label{2.2}
\end{eqnarray}%
with $L/R\equiv \frac{\left( 1\mp \gamma _{5}\right) }{2}$, $s=q^{2}$ which
is just the momentum transfer from heavy to light meson. The amplitude given
in Eq. (\ref{2.2}) contains long distance effects encoded in the form
factors and short distance effects that are hidden in Wilson coefficients.
These Wilson coefficients have been computed at next-to-next leading order
(NNLO) in the SM \cite{13}. Specifically for exclusive decays, the effective
coefficient $C_{9}^{eff}$ can be written as
\begin{equation}
C_{9}^{eff}=C_{9}+Y\left( \hat{s}\right)  \label{resonances1}
\end{equation}%
where the perturbatively calculated result of $Y\left( \hat{s}\right) $ is %
\cite{12}
\begin{equation}
Y_{\mbox{pert}}\left( \hat{s}\right) =\left.
\begin{array}{c}
g\left( \hat{m}_{c}\mbox{,}\hat{s}\right) \left(
3C_{1}+C_{2}+3C_{3}+C_{4}+3C_{5}+C_{6}\right) \\
-\frac{1}{2}g\left( 1\mbox{,}\hat{s}\right) \left(
4C_{3}+4C_{4}+3C_{5}+C_{6}\right) \\
-\frac{1}{2}g\left( 0\mbox{,}\hat{s}\right) \left( C_{3}+3C_{4}\right) +%
\frac{2}{9}\left( 3C_{3}+C_{4}+3C_{5}+C_{6}\right) .%
\end{array}%
\right.  \label{reson-expr}
\end{equation}%
Here the hat denote the normalization in term of $B$ meson mass. For the
explicit expressions of $g$'s and numerical values of the Wilson
coefficients appearing in Eq. (\ref{reson-expr}) we refer to \cite{12}.

Now the new physics effects manifest themselves in rare $B$ decays in two
different ways, either through new contribution to the Wilson coefficients
or through the new operators in the effective Hamiltonian, which are absent
in the SM. In ACD\ model the new physics comes through the Wilson
coefficients. Buras et al. have computed the above coefficients at NLO in
ACD model including the effects of KK modes \cite{03}; we use these results
to study $B\rightarrow K_{1}\tau ^{+}\tau ^{-}$ decay like the one done in
the literature for $B\rightarrow K^{\ast }\left( K_{1}\right) \mu ^{+}\mu
^{-}$ \cite{06, 07}. As it has already been mentioned that ACD model is the
minimal extension of SM with only one extra dimension and it has no extra
operator other than the SM, therefore, the whole contribution from all the
KK states is in the Wilson coefficients, i.e. now they depend on the
additional ACD parameter, the inverse of compactification radius $R$. At
large value of $1/R$ the SM\ phenomenology should be recovered, since the
new states, being more and more massive, decoupled from the low-energy
theory.

Now the modified Wilson coefficients in ACD model contain the contribution
from new particles which are not present in the SM and comes as an
intermediate state in penguin and box diagrams. Thus, these coefficients can
be expressed in terms of the functions $F\left( x_{t},1/R\right) $, $x_{t}=%
\frac{m_{t}^{2}}{M_{W}^{2}}$, which generalize the corresponding SM\
function $F_{0}\left( x_{t}\right) $ according to:
\begin{equation}
F\left( x_{t},1/R\right) =F_{0}\left( x_{t}\right) +\sum_{n=1}^{\infty
}F_{n}\left( x_{t},x_{n}\right)  \label{f-expression}
\end{equation}%
with $x_{n}=\frac{m_{n}^{2}}{M_{W}^{2}}$ and $m_{n}=\frac{n}{R}$ \cite{06}.
The relevant diagrams are $Z^{0}$ penguins, $\gamma $ penguins, gluon
penguins, $\gamma $ magnetic penguins, Chormomagnetic penguins$\ $and the
corresponding functions are $C\left( x_{t},1/R\right) $, $D\left(
x_{t},1/R\right) $, $E\left( x_{t},1/R\right) $, $D^{\prime }\left(
x_{t},1/R\right) $ and $E^{\prime }\left( x_{t},1/R\right) $ respectively.
These functions can be found in \cite{03} and can be summarized as:

$\bullet C_{7}$

In place of $C_{7},$ one defines an effective coefficient $C_{7}^{(0)eff}$
which is renormalization scheme independent \cite{12}:
\begin{equation}
C_{7}^{(0)eff}(\mu _{b})=\eta ^{\frac{16}{23}}C_{7}^{(0)}(\mu _{W})+\frac{8}{%
3}(\eta ^{\frac{14}{23}}-\eta ^{\frac{16}{23}})C_{8}^{(0)}(\mu
_{W})+C_{2}^{(0)}(\mu _{W})\sum_{i=1}^{8}h_{i}\eta ^{\alpha _{i}}
\label{wilson1}
\end{equation}%
where $\eta =\frac{\alpha _{s}(\mu _{W})}{\alpha _{s}(\mu _{b})}$, and
\begin{equation}
C_{2}^{(0)}(\mu _{W})=1,\mbox{ }C_{7}^{(0)}(\mu _{W})=-\frac{1}{2}D^{\prime
}(x_{t},\frac{1}{R}),\mbox{
}C_{8}^{(0)}(\mu _{W})=-\frac{1}{2}E^{\prime }(x_{t},\frac{1}{R});
\label{wilson2}
\end{equation}%
the superscript $(0)$ stays for leading logarithm approximation.
Furthermore:
\begin{eqnarray}
\alpha _{1} &=&\frac{14}{23}\mbox{ \quad }\alpha _{2}=\frac{16}{23}%
\mbox{
\quad }\alpha _{3}=\frac{6}{23}\mbox{ \quad
}\alpha _{4}=-\frac{12}{23}  \nonumber \\
\alpha _{5} &=&0.4086\mbox{ \quad }\alpha _{6}=-0.4230\mbox{ \quad
}\alpha _{7}=-0.8994\mbox{ \quad }\alpha _{8}=-0.1456  \nonumber \\
h_{1} &=&2.996\mbox{ \quad }h_{2}=-1.0880\mbox{ \quad }h_{3}=-\frac{3}{7}%
\mbox{ \quad }h_{4}=-\frac{1}{14}  \nonumber \\
h_{5} &=&-0.649\mbox{ \quad }h_{6}=-0.0380\mbox{ \quad
}h_{7}=-0.0185\mbox{ \quad }h_{8}=-0.0057.  \label{wilson3}
\end{eqnarray}%
The functions $D^{\prime }$ and $E^{\prime }$ are given be eq. (\ref{wilson3}%
) with
\begin{equation}
D_{0}^{\prime }(x_{t})=-\frac{(8x_{t}^{3}+5x_{t}^{2}-7x_{t})}{12(1-x_{t})^{3}%
}+\frac{x_{t}^{2}(2-3x_{t})}{2(1-x_{t})^{4}}\ln x_{t}  \label{wilson4}
\end{equation}%
\begin{equation}
E_{0}^{\prime }(x_{t})=-\frac{x_{t}(x_{t}^{2}-5x_{t}-2)}{4(1-x_{t})^{3}}+%
\frac{3x_{t}^{2}}{2(1-x_{t})^{4}}\ln x_{t}  \label{wilson5}
\end{equation}%
\begin{eqnarray}
D_{n}^{\prime }(x_{t},x_{n}) &=&\frac{%
x_{t}(-37+44x_{t}+17x_{t}^{2}+6x_{n}^{2}(10-9x_{t}+3x_{t}^{2})-3x_{n}(21-54x_{t}+17x_{t}^{2}))%
}{36(x_{t}-1)^{3}}  \nonumber \\
&&+\frac{x_{n}(2-7x_{n}+3x_{n}^{2})}{6}\ln \frac{x_{n}}{1+x_{n}}
\label{wilson6} \\
&&-\frac{%
(-2+x_{n}+3x_{t})(x_{t}+3x_{t}^{2}+x_{n}^{2}(3+x_{t})-x_{n})(1+(-10+x_{t})x_{t}))%
}{6(x_{t}-1)^{4}}\ln \frac{x_{n}+x_{t}}{1+x_{n}}  \nonumber
\end{eqnarray}%
\begin{eqnarray}
E_{n}^{\prime }(x_{t},x_{n}) &=&\frac{%
x_{t}(-17-8x_{t}+x_{t}^{2}+3x_{n}(21-6x_{t}+x_{t}^{2})-6x_{n}^{2}(10-9x_{t}+3x_{t}^{2}))%
}{12(x_{t}-1)^{3}}  \nonumber \\
&&+-\frac{1}{2}x_{n}(1+x_{n})(-1+3x_{n})\ln \frac{x_{n}}{1+x_{n}}  \nonumber
\\
&&+\frac{%
(1+x_{n})(x_{t}+3x_{t}^{2}+x_{n}^{2}(3+x_{t})-x_{n}(1+(-10+x_{t})x_{t}))}{%
2(x_{t}-1)^{4}}\ln \frac{x_{n}+x_{t}}{1+x_{n}}  \label{wilson7}
\end{eqnarray}%
Following \cite{03}, one gets the expressions for the sum over $n:$%
\begin{eqnarray}
\sum_{n=1}^{\infty }D_{n}^{\prime }(x_{t},x_{n}) &=&-\frac{%
x_{t}(-37+x_{t}(44+17x_{t}))}{72(x_{t}-1)^{3}}  \nonumber \\
&&+\frac{\pi M_{w}R}{2}[\int_{0}^{1}dy\frac{2y^{\frac{1}{2}}+7y^{\frac{3}{2}%
}+3y^{\frac{5}{2}}}{6}]\coth (\pi M_{w}R\sqrt{y})  \nonumber \\
&&+\frac{(-2+x_{t})x_{t}(1+3x_{t})}{6(x_{t}-1)^{4}}J(R,-\frac{1}{2})
\nonumber \\
&&-\frac{1}{6(x_{t}-1)^{4}}%
[x_{t}(1+3x_{t})-(-2+3x_{t})(1+(-10+x_{t})x_{t})]J(R,\frac{1}{2})  \nonumber
\\
&&+\frac{1}{6(x_{t}-1)^{4}}[(-2+3x_{t})(3+x_{t})-(1+(-10+x_{t})x_{t})]J(R,%
\frac{3}{2})  \nonumber \\
&&-\frac{(3+x_{t})}{6(x_{t}-1)^{4}}J(R,\frac{5}{2})] ,  \label{wilson8}
\end{eqnarray}%
\begin{eqnarray}
\sum_{n=1}^{\infty }E_{n}^{\prime }(x_{t},x_{n}) &=&-\frac{%
x_{t}(-17+(-8+x_{t})x_{t})}{24(x_{t}-1)^{3}}  \nonumber \\
&&+\frac{\pi M_{w}R}{2}[\int_{0}^{1}dy(y^{\frac{1}{2}}+2y^{\frac{3}{2}}-3y^{%
\frac{5}{2}})\coth (\pi M_{w}R\sqrt{y})]  \nonumber \\
&&-\frac{x_{t}(1+3x_{t})}{(x_{t}-1)^{4}}J(R,-\frac{1}{2})  \nonumber \\
&&+\frac{1}{(x_{t}-1)^{4}}[x_{t}(1+3x_{t})-(1+(-10+x_{t})x_{t})]J(R,\frac{1}{%
2})  \nonumber \\
&&-\frac{1}{(x_{t}-1)^{4}}[(3+x_{t})-(1+(-10+x_{t})x_{t})]J(R,\frac{3}{2})
\nonumber \\
&&+\frac{(3+x_{t})}{(x_{t}-1)^{4}}J(R,\frac{5}{2})]  \label{wilson9}
\end{eqnarray}%
where
\begin{equation}
J(R,\alpha )=\int_{0}^{1}dyy^{\alpha }[\coth (\pi M_{w}R\sqrt{y}%
)-x_{t}^{1+\alpha }\coth (\pi m_{t}R\sqrt{y})].  \label{wilson10}
\end{equation}%
$\bullet C_{9}$

In the ACD model and in the NDR scheme one has
\begin{equation}
C_{9}(\mu )=P_{0}^{NDR}+\frac{Y(x_{t},\frac{1}{R})}{\sin ^{2}\theta _{W}}%
-4Z(x_{t},\frac{1}{R})+P_{E}E(x_{t},\frac{1}{R})  \label{wilson11}
\end{equation}%
where $P_{0}^{NDR}=2.60\pm 0.25$ \cite{12} and the last term is numerically
negligible. Besides
\begin{eqnarray}
Y(x_{t},\frac{1}{R}) &=&Y_{0}(x_{t})+\sum_{n=1}^{\infty }C_{n}(x_{t},x_{n})
\nonumber \\
Z(x_{t},\frac{1}{R}) &=&Z_{0}(x_{t})+\sum_{n=1}^{\infty }C_{n}(x_{t},x_{n})
\label{wilson12}
\end{eqnarray}%
with
\begin{eqnarray}
Y_{0}(x_{t}) &=&\frac{x_{t}}{8}[\frac{x_{t}-4}{x_{t}-1}+\frac{3x_{t}}{%
(x_{t}-1)^{2}}\ln x_{t}]  \nonumber \\
Z_{0}(x_{t}) &=&\frac{18x_{t}^{4}-163x_{t}^{3}+259x_{t}^{2}-108x_{t}}{%
144(x_{t}-1)^{3}}  \nonumber \\
&&+[\frac{32x_{t}^{4}-38x_{t}^{3}+15x_{t}^{2}-18x_{t}}{72(x_{t}-1)^{4}}-%
\frac{1}{9}]\ln x_{t}  \label{wilson13}
\end{eqnarray}%
\begin{equation}
C_{n}(x_{t},x_{n})=\frac{x_{t}}{8(x_{t}-1)^{2}}%
[x_{t}^{2}-8x_{t}+7+(3+3x_{t}+7x_{n}-x_{t}x_{n})\ln \frac{x_{t}+x_{n}}{%
1+x_{n}}]  \label{wilson14}
\end{equation}%
and
\begin{equation}
\sum_{n=1}^{\infty }C_{n}(x_{t},x_{n})=\frac{x_{t}(7-x_{t})}{16(x_{t}-1)}-%
\frac{\pi M_{w}Rx_{t}}{16(x_{t}-1)^{2}}[3(1+x_{t})J(R,-\frac{1}{2}%
)+(x_{t}-7)J(R,\frac{1}{2})]  \label{wilson15}
\end{equation}%
$\bullet C_{10}$

$C_{10}$ is $\mu $ independent and is given by
\begin{equation}
C_{10}=-\frac{Y(x_{t},\frac{1}{R})}{\sin ^{2}\theta _{w}}.  \label{wilson16}
\end{equation}%
The normalization scale is fixed to $\mu =\mu _{b}\simeq 5$ GeV.

\section{Decay rate and forward backward asymmetry}

It is well known that Wilson coefficients give the short distance effects
where as the long distance effects involve the matrix elements of the
operators in Eq. (\ref{2.2}) between the $B$ and $K_{1}$ mesons in $%
B\rightarrow K_{1}\tau ^{+}\tau ^{-}$ process. Using standard
parameterization in terms of the form factors we have \cite{15}:
\begin{eqnarray}
\left\langle K_{1}(k,\varepsilon )\left| V_{\mu }\right| B(p)\right\rangle
&=&i\varepsilon _{\mu }^{\ast }\left( M_{B}+M_{K_{1}}\right) V_{1}(s)
\nonumber \\
&&-(p+k)_{\mu }\left( \varepsilon ^{\ast }\cdot q\right) \frac{V_{2}(s)}{%
M_{B}+M_{K_{1}}}  \nonumber \\
&&-q_{\mu }\left( \varepsilon ^{\ast }\cdot q\right) \frac{2M_{K_{1}}}{s}%
\left[ V_{3}(s)-V_{0}(s)\right]  \label{matrix-1} \\
\left\langle K_{1}(k,\varepsilon )\left| A_{\mu }\right| B(p)\right\rangle
&=&\frac{2i\epsilon _{\mu \nu \alpha \beta }}{M_{B}+M_{K_{1}}}\varepsilon
^{\ast \nu }p^{\alpha }k^{\beta }A(s)  \label{matrix2}
\end{eqnarray}%
where $V_{\mu }=\bar{s}\gamma _{\mu }b$ and $A_{\mu }=\bar{s}\gamma _{\mu
}\gamma _{5}b$ are the vector and axial vector currents respectively and $%
\varepsilon _{\mu }^{\ast }$ is the polarization vector for the final state
axial vector meson.

The relationship between different form factors which also ensures that
there is no kinematical singularity in the matrix element at $s=0$ is
\begin{eqnarray}
V_{3}(s) &=&\frac{M_{B}+M_{K_{1}}}{2M_{K_{1}}}V_{1}(s)-\frac{M_{B}-M_{K_{1}}%
}{2M_{K_{1}}}V_{2}(s)  \label{matrix3} \\
V_{3}(0) &=&V_{0}(0).  \label{matrix4}
\end{eqnarray}%
In addition to the above form factors there are also some penguin form
factors which are:
\begin{eqnarray}
\left\langle K_{1}(k,\varepsilon )\left| \bar{s}i\sigma _{\mu \nu }q^{\nu
}b\right| B(p)\right\rangle &=&\left[ \left( M_{B}^{2}-M_{K_{1}}^{2}\right)
\varepsilon _{\mu }^{\ast }-(\varepsilon ^{\ast }\cdot q)(p+k)_{\mu }\right]
F_{2}(s)  \nonumber \\
&&+(\varepsilon ^{\ast }\cdot q)\left[ q_{\mu }-\frac{s}{%
M_{B}^{2}-M_{K_{1}}^{2}}(p+k)_{\mu }\right] F_{3}(s)  \nonumber \\
&&  \label{matrix-5} \\
\left\langle K_{1}(k,\varepsilon )\left| \bar{s}i\sigma _{\mu \nu }q^{\nu
}\gamma _{5}b\right| B(p)\right\rangle &=&-i\epsilon _{\mu \nu \alpha \beta
}\varepsilon ^{\ast \nu }p^{\alpha }k^{\beta }F_{1}(s)  \label{matrix-6}
\end{eqnarray}%
with $F_{1}(0)=2F_{2}(0).$

Form factors are the non-perturbative quantities and are the scalar function
of the square of momentum transfer. Different models are used to calculate
these form factors. The form factors we use here in the analysis of the
observables like decay rate, forward backward asymmetry and polarization
asymmetries of final state $\tau $ in $B\rightarrow K_{1}\tau ^{+}\tau ^{-}$
have been calculated using Ward identities. The detailed calculation and
their expressions are given in ref. \cite{15} and can be summarized as:
\begin{eqnarray}
A\left( s\right) &=&\frac{A\left( 0\right) }{\left( 1-s/M_{B}^{2}\right)
(1-s/M_{B}^{\prime 2})}  \nonumber \\
V_{1}(s) &=&\frac{V_{1}(0)}{\left( 1-s/M_{B_{A}^{\ast }}^{2}\right) \left(
1-s/M_{B_{A}^{\ast }}^{\prime 2}\right) }\left( 1-\frac{s}{%
M_{B}^{2}-M_{K_{1}}^{2}}\right)  \label{form-factors} \\
V_{2}(s) &=&\frac{\tilde{V}_{2}(0)}{\left( 1-s/M_{B_{A}^{\ast }}^{2}\right)
\left( 1-s/M_{B_{A}^{\ast }}^{\prime 2}\right) }-\frac{2M_{K_{1}}}{%
M_{B}-M_{K_{1}}}\frac{V_{0}(0)}{\left( 1-s/M_{B}^{2}\right) \left(
1-s/M_{B}^{\prime 2}\right) }  \nonumber
\end{eqnarray}%
with
\begin{eqnarray}
A(0) &=&-(0.52\pm 0.05)  \nonumber \\
V_{1}(0) &=&-(0.24\pm 0.02)  \nonumber \\
\tilde{V}_{2}(0) &=&-(0.39\pm 0.03).  \label{Num-f-factor}
\end{eqnarray}%
The corresponding values for $B\rightarrow K^{\ast }$ form factors at $s=0$
are given by \cite{07}
\begin{eqnarray}
V(0) &=&(0.29\pm 0.04)  \nonumber \\
A_{1}(0) &=&(0.23\pm 0.03)  \label{Kstformfactor} \\
\tilde{A}_{2}(0) &=&(0.33\pm 0.05).  \nonumber
\end{eqnarray}

\begin{table}[tbp]
\caption{Default value of input parameters used in the calculation}
\label{table1}
\begin{center}
\begin{tabular}{ll}
\hline
$m_{W}$ & $80.41$ GeV \\
$m_{Z}$ & $91.1867$ GeV \\
$sin^{2}\theta _{W}$ & $0.2233$ \\
$m_{c}$ & $1.4$ GeV \\
$m_{b,pole}$ & $4.8\pm 0.2$ GeV \\
$m_{t}$ & $173.8\pm 5.0$ GeV \\
$\alpha _{s}\left( m_{Z}\right) $ & $0.119\pm 0.0058$ \\
$f_{B}$ & $\left( 200\pm 30\right) $ MeV \\
$\left| V_{ts}^{*}V_{tb}\right| $ & $0.0385$ \\ \hline
\end{tabular}%
\end{center}
\end{table}

Following the notation from ref. \cite{09a}, the differential decay rate in
terms of the auxiliary functions can be written as%
\begin{equation}
\frac{d\Gamma }{ds}=\frac{G_{F}^{2}\left| V_{tb}V_{ts}^{\ast }\right|
^{2}\alpha ^{2}}{2^{11}\pi ^{5}}\frac{\lambda ^{1/2}\left(
M_{B}^{2},M_{K_{1}}^{2},s\right) }{3M_{B}^{3}M_{K_{1}}^{2}s}\sqrt{1-\frac{%
4m_{\tau }^{2}}{s}}g\left( s\right)  \label{ddrate}
\end{equation}%
where $\lambda \left( a,b,c\right) =a^{2}+b^{2}+c^{2}-2ab-2bc-2ca$ and the
function $g\left( s\right) $ is:%
\begin{eqnarray}
g\left( s\right) &=&24\left| D_{0}\right| ^{2}m_{\tau
}^{2}M_{K_{1}}^{2}\lambda +8sM_{K_{1}}^{2}\lambda \left[ \left( 2m_{\tau
}^{2}+s\right) \left| A\right| ^{2}-\left( 4m_{\tau }^{2}-s\right) \left|
C\right| ^{2}\right]  \nonumber \\
&&+\lambda \left[ \left( 2m_{\tau }^{2}+s\right) \left| B_{1}+\left(
M_{B}^{2}-M_{K_{1}}^{2}-s\right) B_{2}\right| ^{2}-\left( 4m_{\tau
}^{2}-s\right) \left|
\begin{array}{c}
D_{1} \\
+\left( M_{B}^{2}-M_{K_{1}}^{2}-s\right) D_{2}%
\end{array}%
\right| ^{2}\right]  \nonumber \\
&&+4sM_{K_{1}}^{2}\left[ \left( 2m_{\tau }^{2}+s\right) \left( 3\left|
B_{1}\right| ^{2}-\lambda \left| B_{2}\right| ^{2}\right) -\left( 4m_{\tau
}^{2}-s\right) \left( 3\left| D_{1}\right| ^{2}-\lambda \left| D_{2}\right|
^{2}\right) \right] .  \label{expressiong}
\end{eqnarray}%
The auxiliary functions contain the short distance contribution (Wilson
coefficients) as well as the long distance contribution (form factors):
\begin{eqnarray}
A &=&4\left( m_{b}+m_{s}\right) \frac{C_{7}^{eff}}{s}F_{1}(s)-\frac{A_{0}(s)%
}{M_{B}+M_{K_{1}}}C_{9}^{eff}(s)  \nonumber \\
B_{1} &=&\left( M_{B}+M_{K_{1}}\right) \left[ C_{9}^{eff}(s)V_{1}(s)+\frac{%
4m_{b}}{s}C_{7}^{eff}\left( M_{B}-M_{K_{1}}\right) F_{2}\left( s\right) %
\right]  \nonumber \\
B_{2} &=&-\left[ \frac{4m_{b}}{s}C_{7}^{eff}\left( F_{2}\left( s\right) +s%
\frac{F_{3}\left( s\right) }{M_{B}^{2}-M_{K_{1}}^{2}}\right) +C_{9}^{eff}(s)%
\frac{V_{2}(s)}{M_{B}+M_{K_{1}}}\right]  \nonumber \\
C &=&-C_{10}\frac{A(s)}{M_{B}+M_{K_{1}}}  \nonumber \\
D_{0} &=&C_{10}V_{0}\left( s\right)  \nonumber \\
D_{1} &=&C_{10}V_{1}\left( s\right) \left( M_{B}+M_{K_{1}}\right)  \nonumber
\\
D_{2} &=&C_{10}\frac{V_{2}\left( s\right) }{M_{B}+M_{K_{1}}}.
\label{afunctions1}
\end{eqnarray}%
Thus, integrating Eq. (\ref{ddrate}) on $s$ and using the value of the form
factors defined in Eq. (\ref{form-factors}), the numerical value of the
branching ratio $B\rightarrow K_{1}\tau ^{+}\tau ^{-}$ is
\[
{B}\left( B\rightarrow K_{1}\tau ^{+}\tau ^{-}\right) =\left( \,0.6\pm
0.1\right) \times 10^{-7}.
\]
The error in the value reflects the uncertainty from the form factors, and
due to the variation of input parameters like CKM matrix elements, decay
constant of $B$ meson and masses as defined in Table~\ref{table1}.

\begin{figure}[tbp]
\includegraphics[scale=0.8]{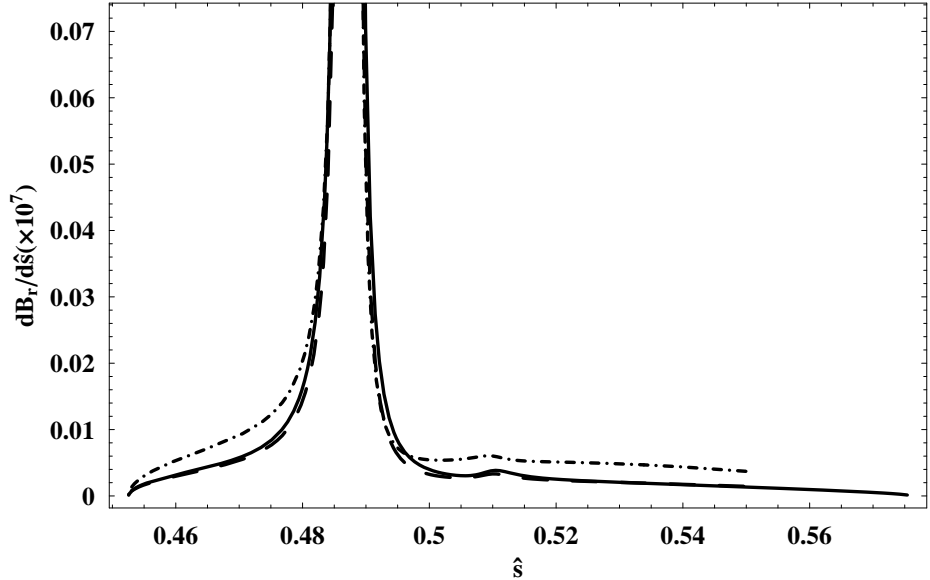}
\caption{The differential branching ratio as a function of $\hat{s}$ is
plotted using the form factors defined in Eq. (\ref{form-factors}). The
solid line denotes the SM result, dashed-dotted line is for $1/R=200$ GeV
and dashed line is for $1/R=500$ GeV. All the input parameters are taken at
their central values.}
\end{figure}

It is already mentioned that in ACD model there is no new opertor beyond the
SM and new physics will come only through the Wilson coefficients. To see
this, the differential branching ratio against $\hat{s}\left(
=s/M_{B}^{2}\right) $ is plotted in Fig. 1 using the central values of input
parameters. One can see that the effect of KK contribution in the Wilson
coefficient are modest for $1/R=200$ GeV at low value of $\hat{s}$ but such
effects are obscured by the uncertainties involved in different parameters
like, form factors, CKM matrix elements, etc at large value of $\hat{s}$.

Another observable is the forward backward asymmetry ($\mathcal{A}_{\mbox{FB}%
}$), which is also very useful tool for looking new physics. It has been
shown by Ishtiaq \textit{et al.} \cite{07} that zero of the forward backward
asymmetry is considerably shifted to the left in ACD model for $B\rightarrow
K_{1}\mu ^{+}\mu ^{-}$. What we have shown in Fig.~2 is the differential
forward backward asymmetry with $\hat{s}$ for $B\rightarrow K_{1}\tau
^{+}\tau ^{-}$. Again the sensitivity of the zero on the extra dimension is
very mild for $1/R=200$ GeV.

\begin{figure}[tbp]
\vspace{-1cm} \includegraphics[scale=0.6]{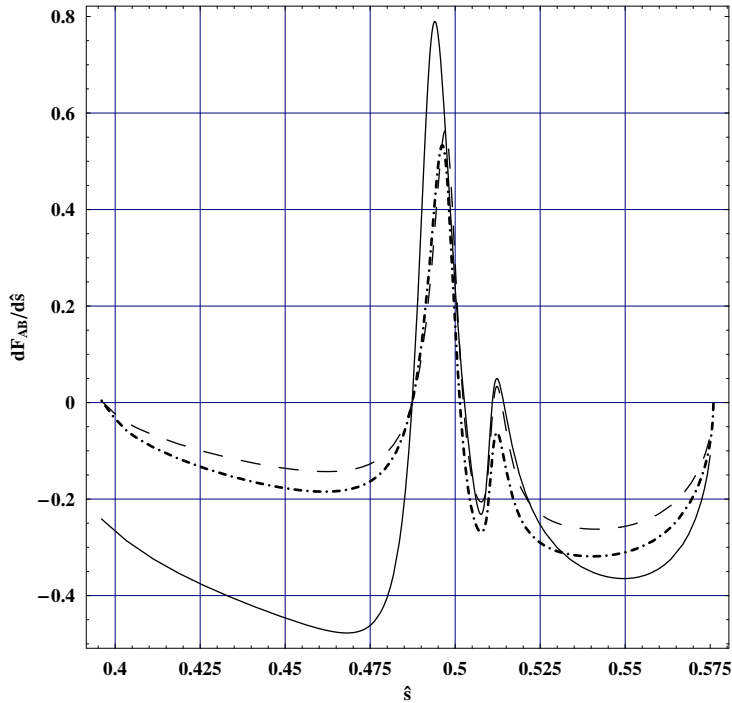}
\caption{The differential forward-backward (FB) asymmetry as a function of $
\hat{s}$ is plotted using the form factors defined in Eq. (\ref{form-factors}%
). The solid line denotes the SM result, dashed line is for $1/R=200$ GeV
and dashed-dotted line is for $1/R=500$ GeV. All the input parameters are
taken at their central values.}
\end{figure}

\section{Polarization asymmetries of final state leptons}

In this section we will discuss the final state lepton polarization
asymmetries by following the notation defined in ref. \cite{09a}. To compute
these for $B$ decays in $\tau $ leptons we consider the spin vector $n$ of $%
\tau ^{-}$, with $n^{2}=-1$ and $k_{1}\cdot n=0$, $k_{1}$ is the momentum of
$\tau ^{-}$. Now in the rest frame of the $\tau ^{-}$ lepton, one can define
the three orthogonal unit vectors: $e_{L}$, $e_{N}$ and $e_{T}$
corresponding to the longitudinal $n_{L}$, normal $n_{N}$ and transverse $%
n_{T}$ polarization vectors:%
\begin{eqnarray}
n_{L} &=&\left( 0,e_{L}\right) =\left( 0,\frac{\vec{k}_{1}}{\left| \vec{k}%
_{1}\right| }\right)  \nonumber \\
n_{N} &=&\left( 0,e_{N}\right) =\left( 0,\frac{\vec{p}\times \vec{k}_{1}}{%
\left| \vec{p}\times \vec{k}_{1}\right| }\right)  \label{definations} \\
n_{T} &=&\left( 0,e_{T}\right) =\left( 0,e_{N}\times e_{T}\right)  \nonumber
\end{eqnarray}%
where, $\vec{p}$ and $\vec{k}_{1}$ are the three momenta of $K_{1}$ and $%
\tau ^{-}$ in the rest frame of the lepton pair. If we choose the $z$-axis
as the momentum direction of $\tau ^{-}$ in the rest frame of lepton pair,
then $k_{1}=\left( E_{1},0,0,\left| \vec{k}_{1}\right| \right) $. Now
boosting the spin vector $n$ defined in Eq. (\ref{definations}) in the rest
frame of lepton pair, the normal and transverse vectors $n_{N}$, $n_{T}$
remains unchanged but the longitudinal polarization vector changes. Their
new form becomes%
\begin{eqnarray}
n_{N} &=&(0,1,0,0)  \nonumber \\
n_{T} &=&\left( 0,0,-1,0\right)  \label{boost} \\
n_{L} &=&\frac{1}{m_{\tau }}\left( \left| \vec{k}_{1}\right|
,0,0,E_{1}\right) .  \nonumber
\end{eqnarray}%
The polarization asymmetry for negatively charged lepton $\tau ^{-}$ for
each value of the square of momentum transfer to the lepton pairs, $s$ can
be defined as:%
\begin{equation}
\mathcal{A}_{i}\left( s\right) =\frac{\frac{d\Gamma }{ds}\left( n_{i}\right)
-\frac{d\Gamma }{ds}\left( -n_{i}\right) }{\frac{d\Gamma }{ds}\left(
n_{i}\right) +\frac{d\Gamma }{ds}\left( -n_{i}\right) }  \label{asymmetry}
\end{equation}%
with $i=L,\,T$ and $N$.

Thus, for $B\rightarrow K_{1}\tau ^{+}\tau ^{-}$ the expression of the
longitudinal $\mathcal{A}_{L}\left( s\right) $ and transverse $\mathcal{A}%
_{T}\left( s\right) $ polarization asymmetries of $\tau ^{-}$ becomes \cite%
{09a}:%
\begin{eqnarray}
\mathcal{A}_{L}\left( s\right) &=&2s\sqrt{1-\frac{4m_{\tau }^{2}}{s}}\frac{1%
}{g\left( s\right) }\bigg\{8sM_{K_{1}}^{2}\mbox{Re}[B_{1}D_{1}^{\ast
}+\lambda AC^{\ast }]  \label{Longti} \\
&&+\mbox{Re}\left[ \left( M_{B}^{2}-M_{K_{1}}^{2}-s\right) B_{1}+\lambda
B_{2}\right] \left[ \left( M_{B}^{2}-M_{K_{1}}^{2}-s\right) D_{1}^{\ast
}+\lambda D_{2}^{\ast }\right] \bigg\}  \nonumber
\end{eqnarray}%
\begin{equation}
\mathcal{A}_{T}\left( s\right) =3\pi m_{\tau }M_{K_{1}}\frac{\lambda \sqrt{s}%
}{g\left( s\right) }\left\{ -4\mbox{Re}[AB_{1}^{\ast }]M_{K_{1}}s+\mbox{Re}%
\left[ D_{0}B_{1}^{\ast }\left( M_{B}^{2}-M_{K_{1}}^{2}-s\right) +\lambda
D_{0}B_{2}^{\ast }\right] \right\}  \label{Transverse}
\end{equation}%
with $\lambda =\lambda \left( M_{B}^{2},M_{K_{1}}^{2},s\right) $. Now, while
calculating these asymmetries we do not consider the contribution associated
with the real $c\bar{c}$ resonances in $C_{9}^{eff}$, as these can be
removed by using an appropriate kinematical cuts \cite{09a}. It is clear
from Eq. (\ref{Longti}) that the value of longitudinal polarization
asymmetry vanishes when $s=4m_{\tau }^{2}$. In Fig. 3 we have shown the
effect of extra dimension on the value of the asymmetries.\ One can see that
longitudinal polarization has the largest value at large momentum transfer
(large value of $\hat{s}$) and is least sensitive to the compactification
radius $1/R$. The effects of extra dimension are move evident for the
transverse polarization whose value decreases with the decrease of $1/R$
down to $1/R=200$ and the change is maximum for low value of $\hat{s}$.

\begin{figure}[tbp]
\includegraphics[scale=0.8]{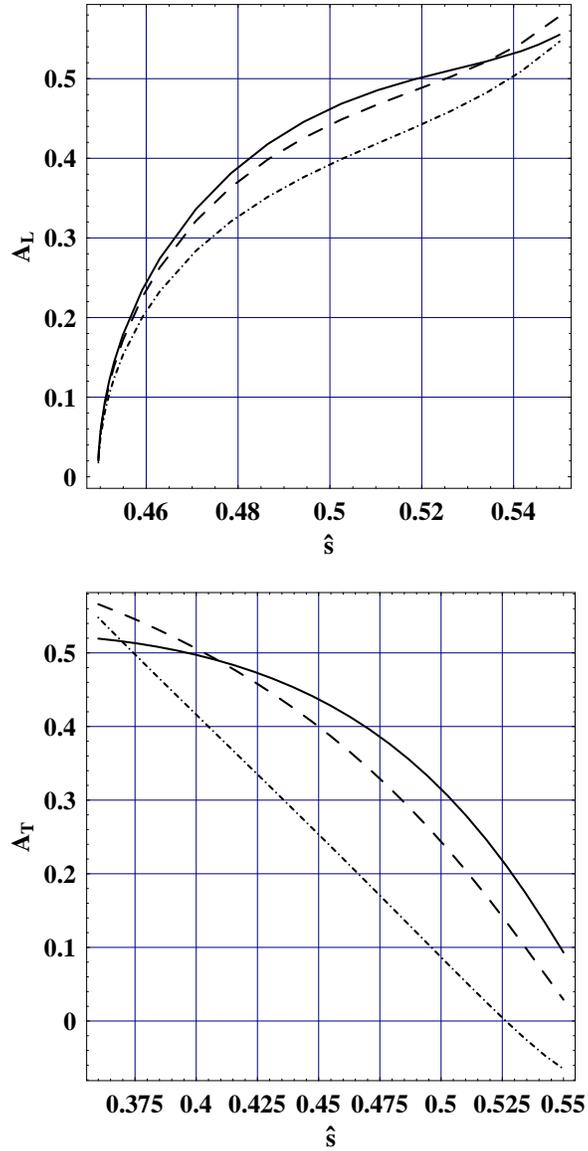}
\caption{ Longitudinal (upper panel) and transverse (lower panel) $\protect%
\tau ^{-}$ polarization asymmetry in $B\rightarrow K_{1}\protect\tau ^{-}%
\protect\tau ^{+}$ is plotted as a function of $\hat{s}$ using form factor
defined in Eq. (\ref{form-factors}). The solid line denotes the SM result,
dashed line is for $1/R=200$ GeV and long-dashed line is for $1/R=500$ GeV.
All the input parameters are taken at their central values.}
\end{figure}

\section{Helicity Fractions of $K_{1}$ in $B\rightarrow K_{1}\ell ^{+}\ell
^{-}$}

In this section, we study the helicity fractions of the $K_{1}$ produced in
the final state, which is another interesting variable. For $K^{\ast }$
meson, the longitudinal helicity fraction $f_{L}$ in the modes $B\rightarrow
K^{\ast }\ell ^{+}\ell ^{-}$ $\left( \ell =e,\,\mu \right) $, has been
measured by Babar Collaboration in two bins of momentum transfer \cite{16}.
The results are:%
\begin{eqnarray}
f_{L} &=&0.77_{-0.30}^{+0.63}\pm 0.07\,\ \ \ \ \ \ \ 0.1\leq s\leq 8.41%
\mbox{
GeV}^{2}  \nonumber \\
f_{L} &=&0.51_{-0.25}^{+0.22}\pm 0.08\,\ \ \ \ \ \ \ s\geq 10.24\mbox{ GeV}%
^{2}  \label{flexperimental1}
\end{eqnarray}%
while the average value of $f_{L}$ in the full $s$ range is \cite{09a}%
\begin{equation}
f_{L}=0.63_{-0.19}^{+0.18}\pm 0.05\,\ \ \ \ \ \ \ s\geq 0.1\mbox{
GeV}^{2}.  \label{flaverage}
\end{equation}%
The expressions of $B\rightarrow K^{\ast }\ell ^{+}\ell ^{-}$ differential
decays widths with $K^{\ast }$ longitudinal ($L$) or transversely ($\pm $)
polarized are calculated by Colangelo et al. \cite{09a}. We will translate
the same results for $B\rightarrow K_{1}\ell ^{+}\ell ^{-}$ as $K^{\ast }$
and $K_{1}$ differ by $\gamma _{5}$ in their distribution amplitudes. The
result reads as follows:
\begin{eqnarray}
\frac{d\Gamma _{L}\left( s\right) }{ds} &=&\frac{G_{F}^{2}\left|
V_{tb}V_{ts}^{\ast }\right| ^{2}\alpha ^{2}}{2^{11}\pi ^{5}}\frac{\lambda
^{1/2}\left( M_{B}^{2},M_{K_{1}}^{2},s\right) }{M_{B}^{3}}\sqrt{1-\frac{%
4m_{\ell }^{2}}{s}}\frac{1}{3}A_{L}  \nonumber \\
\frac{d\Gamma _{+}\left( s\right) }{ds} &=&\frac{G_{F}^{2}\left|
V_{tb}V_{ts}^{\ast }\right| ^{2}\alpha ^{2}}{2^{11}\pi ^{5}}\frac{\lambda
^{1/2}\left( M_{B}^{2},M_{K_{1}}^{2},s\right) }{M_{B}^{3}}\sqrt{1-\frac{%
4m_{\ell }^{2}}{s}}\frac{4}{3}A_{+}  \nonumber \\
\frac{d\Gamma _{-}\left( s\right) }{ds} &=&\frac{G_{F}^{2}\left|
V_{tb}V_{ts}^{\ast }\right| ^{2}\alpha ^{2}}{2^{11}\pi ^{5}}\frac{\lambda
^{1/2}\left( M_{B}^{2},M_{K_{1}}^{2},s\right) }{M_{B}^{3}}\sqrt{1-\frac{%
4m_{\ell }^{2}}{s}}\frac{4}{3}A_{-}  \label{helicities}
\end{eqnarray}%
with%
\begin{eqnarray}
A_{L} &=&\frac{1}{sM_{K_{1}}^{2}}\{24\left| D_{0}\right| ^{2}m_{\ell
}^{2}M_{K_{1}}^{2}\lambda +\left( 2m_{\ell }^{2}+s\right) \left| \left(
M_{B}^{2}-M_{K_{1}}^{2}-s\right) B_{1}+\lambda B_{2}\right| ^{2}  \nonumber
\\
&&+\left( s-4m_{\ell }^{2}\right) \left| \left(
M_{B}^{2}-M_{K_{1}}^{2}-s\right) D_{1}+\lambda D_{2}\right| ^{2}\}
\label{al1}
\end{eqnarray}%
and%
\begin{eqnarray}
A_{-} &=&\left( s-4m_{\ell }^{2}\right) \left| D_{1}+\lambda ^{1/2}C\right|
^{2}+\left( s+2m_{\ell }^{2}\right) \left| B_{1}+\lambda ^{1/2}A\right| ^{2}
\nonumber \\
A_{+} &=&\left( s-4m_{\ell }^{2}\right) \left| D_{1}-\lambda ^{1/2}C\right|
^{2}+\left( s+2m_{\ell }^{2}\right) \left| B_{1}-\lambda ^{1/2}A\right| ^{2}.
\label{aplusminu}
\end{eqnarray}%
The auxiliary functions and the corresponding form factors are defined in
Eqs. (\ref{afunctions1}) and (\ref{Num-f-factor}). The various helicity
amplitudes are defined as \cite{09a}:%
\begin{eqnarray}
f_{L}\left( s\right) &=&\frac{d\Gamma _{L}\left( s\right) /ds}{d\Gamma
\left( s\right) /ds}  \nonumber \\
f_{\pm }\left( s\right) &=&\frac{d\Gamma _{\pm }\left( s\right) /ds}{d\Gamma
\left( s\right) /ds}  \label{fractions} \\
f_{T}\left( s\right) &=&f_{+}\left( s\right) +f_{-}\left( s\right).
\nonumber
\end{eqnarray}%
The helicity fractions for $K^{\ast }$ has been considered in SM and some of
its extensions \cite{09a, 17}. In Fig.~4 we have shown the results of the
helicity fractions of $K^{\ast }$ using the central value of the form
factors and other parameters defined in ref. \cite{07} in SM\ and for two
values of the compactification radius $1/R$. The lepton in the final state
is considered to be $e$ or $\mu $. The effect of extra dimensions are very
mild for the low value of momentum transfer. One can see that the value of
the longitudinal helicity agrees with the experimental data within the
experimental uncertainties both for the small and large value of momentum
transfer. Thus, measurement of transverse helicity fraction will
discriminate between the different models \cite{09a}.

\begin{figure}[tbp]
\includegraphics[scale=0.8]{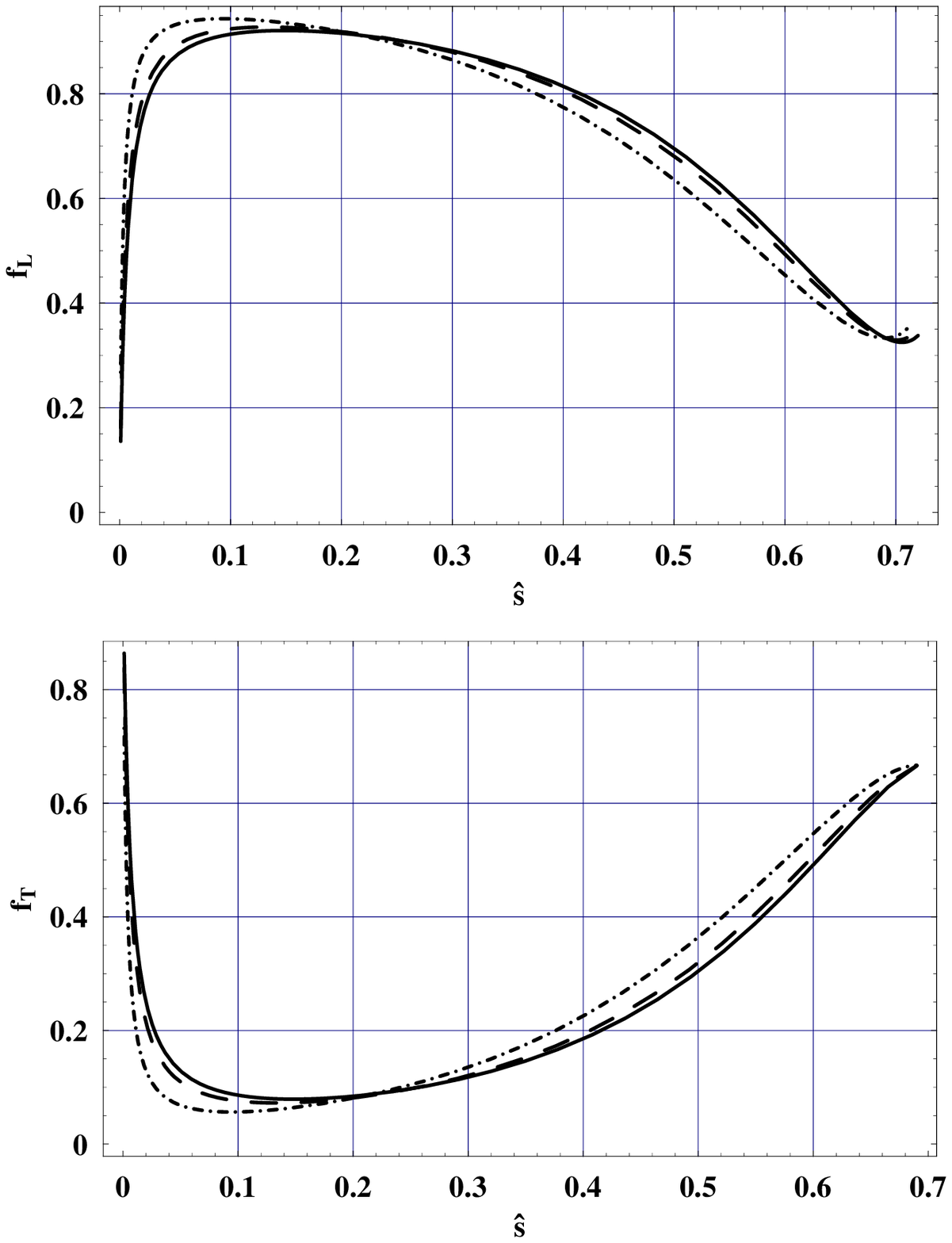}
\caption{: Longitudinal (upper panel) and transverse (lower panel) $K^{\ast
} $ helicity fractions in $B\rightarrow K^{\ast }\ell ^{-}\ell ^{+}\left(
\ell =e,\,\protect\mu \right) $ are obtained using form factor defined in
Eq. (\ref{Kstformfactor}). The solid line denotes the SM result, dashed line
is for $1/R=200$ GeV and long-dashed line is for $1/R=500$ GeV. All the
input parameters are taken at their central values.}
\end{figure}

The results for the case of $K_{1}$ are shown in Fig. 5 and Fig. 6 in SM and
in UED\ model for two values of $1/R$. Fig. 5 shows the helicity fractions
of $K_{1}$ when we considered the $e$ and $\mu $ as the final state lepton
in $B\rightarrow $ $K_{1}\ell ^{+}\ell ^{-}$ and take all the input
parameters at their central values. One can see that the effect of extra
dimension are very prominent at the small value of momentum transfer. These
effects are construtcive for the case of transverse helicity fraction and
destructive for the longitudinal one. Similarly, Fig. 6 dipicts the results
when we have considered the $\tau $ in the final state. Again, the effects
of extra dimension are very modest at the small value of momentum transfers $%
\hat{s}$ where $f_{L}$ is maximum and $f_{T}$ is minimum. From these two
figures it is also clear that at each value of momentum transfer, $%
f_{L}\left( \hat{s}\right) +f_{T}\left( \hat{s}\right) =1$. Thus we can say
that the measurement of helicity fractions of $K_{1}$ will be possible in
future $B$ factories.

\begin{figure}[tbp]
\includegraphics[scale=0.6]{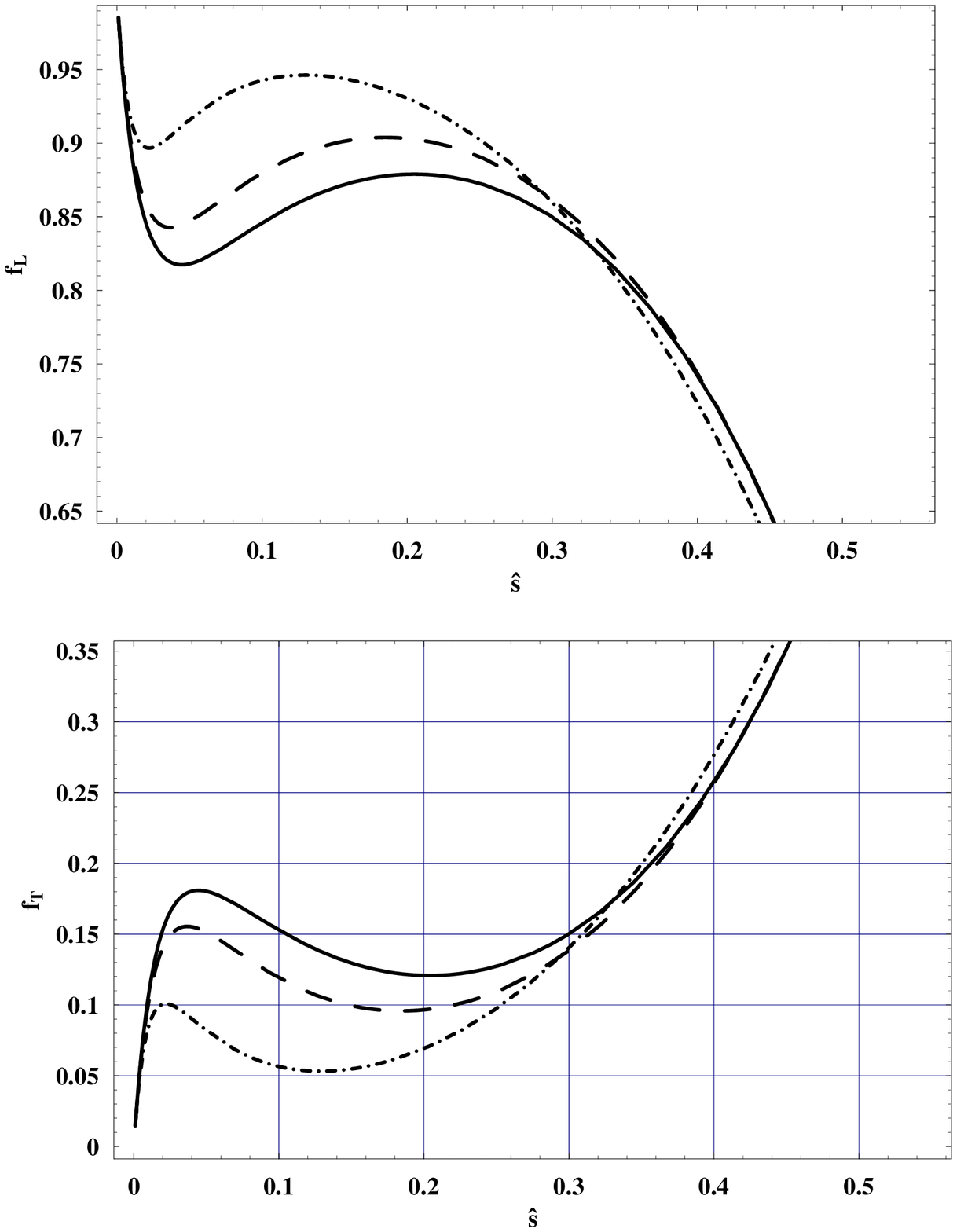}
\caption{ Longitudinal (upper panel) and transverse (lower panel) $K_{1}$
helicity fractions in $B\rightarrow K_{1}\ell ^{-}\ell ^{+}\left( \ell =e,\,%
\protect\mu \right) $ are obtained using form factor defined in Eq. (\ref%
{form-factors}). The solid line denotes the SM result, dashed line is for $%
1/R=200$ GeV and long-dashed line is for $1/R=500$ GeV. All the input
parameters are taken at their central values.}
\end{figure}

\begin{figure}[tbp]
\includegraphics[scale=0.8]{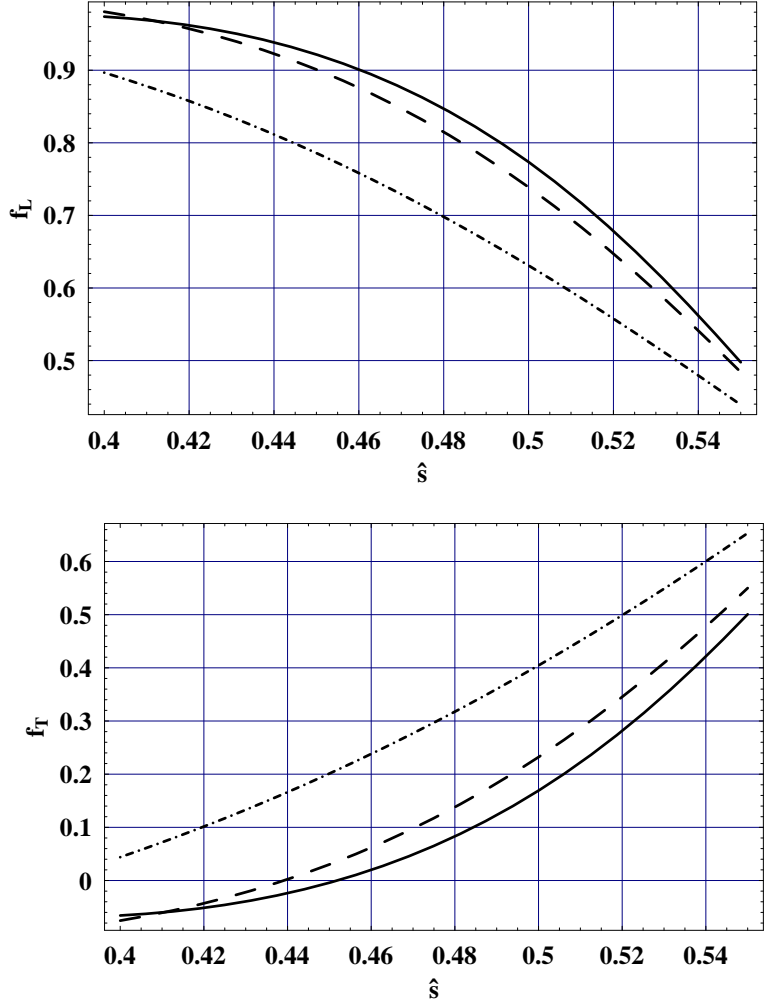}
\caption{Longitudinal (upper panel) and transverse (lower panel) $K_{1}$
helicity fractions in $B\rightarrow K_{1}\protect\tau ^{-}\protect\tau ^{+}$
are obtained using form factor defined in Eq. (\ref{form-factors}). The
solid line denotes the SM result, dashed line is for $1/R=200$ GeV and
long-dashed line is for $1/R=500$ GeV. All the input parameters are taken at
their central values.}
\end{figure}

\section{Conclusion}

In this paper, we have analyzed the spin effects in semileptonic decay $%
B\rightarrow K_{1}\tau ^{+}\tau ^{-}$ both in SM and in ACD model, which is
minimal extension of SM with only one extra dimension. We studied the
dependence of the physical observables like decay rate, forward backward
asymmetry and polarization asymmetry on the inverse of compactification
radius $1/R$. The effects of extra dimension to these observables are very
mild, but still observable. Among the polarization asymmetries the most
sensitive is the transverse one, where the effects of extra dimension for $%
1/R=200$ GeV are very clear in the low momentum transfer range. $K^{\ast }$
helicity fractions, for which some results for $e$ and $\mu $ in the final
state are already available have also been discussed and compared with the
existing results in the literature. Finally, following the same analogy we
considered the $K_{1}$ helicity fractions both in SM and in ACD model. The
future experiments, where more data is expected, will put stringent
constraints on the compactification radius and also give us some deep
understanding of $B$-physics and take us step forward towards the ultimate
theory of fundamental interactions.

\section*{Acknowledgements}

This work of Jamil and Lu is partly supported by National Science Foundation
of China under the Grant Numbers 10735080 and 10625525. One of us (Asif)
would like to thank\ Riazuddin, Fayyazuddin, Ishtiaq and Ali for useful
discussions.


\begin{thebibliography}{99}
\bibitem{01} I. Antoniadis, Phys. Lett. \textbf{B246}, 377 (1990); K. R.
Dienes, E. Dudas and T. Gherghetta, Phys. Lett. \textbf{B436}, 55 (1998); N.
Arkani-Hamed and M. Schmaltz, Phys. Rev. \textbf{D61}, 033005 (2000); N.
Arkani-Hamed, S. Dimopoulos and G. R. Dvali, Phys. Lett. \textbf{B429}, 263
(1998); L. Randall and R. Sundrum, Phys. Rev. Lett. \textbf{83}, 3370
(1999); L. Randall and R. Sundrum, Phys. Rev. Lett. \textbf{83}, 4690 (1999)

\bibitem{02} T. Appelquist, H. C. Cheng and B. A. Dobrescu, Phys. Rev.
\textbf{D64}, 035002 (2001).

\bibitem{03} A. J. Buras, M. Spranger and A. Weiler, Nucl. Phys. \textbf{B660%
}, 225 (2003); A. J. Buras, A. Poschenrieder, M. Spranger and A. Weiler,
Nucl. Phys. \textbf{B678}, 455 (2004).

\bibitem{03a} S. L. Glashow, J. Iliopoulos, and L. Maiani, Phys. Rev.
\textbf{D2}, 1285 (1970).

\bibitem{04} K. Agashe, N. G. Deshpande and G. H. Wu, Phys. Lett. \textbf{%
B514}, 309 (2001).

\bibitem{05} T. Appelquist and H. U. Yee, Phys. Rev. \textbf{D67}, 055002
(2003).

\bibitem{06} P. Colangelo, F. De Fazio, R. Ferrandes and T. N. Pham, Phys.
Rev. \textbf{D73}, 115006 (2006).

\bibitem{07} Ishtiaq Ahmed, M. Ali Paracha, M. Jamil\ Aslam, arXiv:
hep-ph/0802.0740

\bibitem{08} R. Mohanta and A. K. Giri, Phys.Rev.\textbf{D75}, 035008
(2007), arXiv: hep-ph/0611068.

\bibitem{09} T. M. Aliev and M. Savci, arXiv:hep-ph/0606225; T. M. Aliev, M.
Savci and B. B. Sirvanli, arXiv:hep-ph/0608143.

\bibitem{09a} P. Colangelo, F. De Fazio, R. Ferrandes, T. H. Pham, Phys.Rev.%
\textbf{D74},115006 (2006), arXiv: hep-ph/0610044.

\bibitem{10} J. L. Hewett, Phys. Rev. \textbf{D53}, 4964 (1996).

\bibitem{11} A. Ali, P. Ball, L. T. Handoko and G. Hiller, Phys. Rev.
\textbf{D61}, 074024 (2000) [arXiv:hep-ph/9910221].

\bibitem{12} A. J.\ Buras \textit{et al}., Nucl. Phys. \textbf{B424}, 374
(1994).

\bibitem{13} C. Bobeth, M. Misiak and J. Urban, Nucl. Phys. \textbf{B574},
291 (2000); H. H. Asatrian, H. M. Asatrian, C. Greub andM.Walker, Phys.
Lett. \textbf{B507}, 162 (2001); Phys. Rev. \textbf{D65}, 074004 (2002);
Phys. Rev. \textbf{D66}, 034009 (2002); H. M. Asatrian, K. Bieri, C. Greub
and A. Hovhannisyan, Phys. Rev. \textbf{D66}, 094013 (2002); A. Ghinculov,
T. Hurth, G. Isidori and Y. P. Yao, Nucl. Phys. \textbf{B648}, 254 (2003);
A. Ghinculov, T. Hurth, G. Isidori and Y. P. Yao, Nucl. Phys. \textbf{B685},
351 (2004); C. Bobeth, P. Gambino, M. Gorbahn and U. Haisch, JHEP \textbf{%
0404}, 071 (2004).

\bibitem{15} M.\ Ali Paracha, Ishtiaq Ahmed and M. Jamil\ Aslam, Eur. Phys.
J.\textbf{C52}, 967-973 (2007)

\bibitem{16} B. Aubert et al. [BABAR Collaboration], Phys. Rev. \textbf{D73}%
, 092001 (2006).

\bibitem{17} T. M. Aliev, A. Ozpineci and M. Savci, Phys. Lett. \textbf{B511}%
, 49 (2001).
\end{thebibliography}
\end{document}